\newcounter{myctr}
\def\myitem{\refstepcounter{myctr}\bibfont\noindent\ifnum\themyctr>9\else
\phantom{0}\fi\hangindent17pt\themyctr.\enskip}

\documentclass{ws-ijqi}

\usepackage{bbm}

\begin{document}

\markboth{Argentieri, Benatti, Floreanini, Marzolino}
{Entanglement and Squeezing with Identical Particles}

\catchline{}{}{}{}{}

\title{ENTANGLED IDENTICAL PARTICLES AND NOISE}

\author{G. ARGENTIERI$^{\dagger,\star}$, F. BENATTI$^{\dagger,\star}$, R. FLOREANINI$^\star$ and U. MARZOLINO$^{\ast}$}

\address{${}^\dagger$Dipartimento di Fisica, Universit\`a di Trieste,
Trieste, Italy\\
${}^\star$Istituto Nazionale di Fisica Nucleare, Sezione di Trieste,
Trieste, Italy\\
${}^{\ast}$Universit\`a di Salerno, Salerno, Italy}

\maketitle


\begin{abstract}
For systems of identical Bosons, it is necessary to reformulate the notions of separability and entanglement in algebraic terms shifting the emphasis from the particle aspect of first quantization to the mode description typical of second quantization. Within this new framework, we show that, unlike for systems consisting of distinguishable qubits, negativity is an exhaustive bipartite entanglement witness for systems with fixed number of Bosons; further, we investigate the impact of dephasing noise in relation to the use of such many-body Bosonic systems in metrological applications.
\end{abstract}


\section{Introduction}

No longer an epistemological curiosity, entanglement is becoming an experimentally accessible resource; in particular, entangled $N$-qubit states have been proposed as means to beat the so-called shot-noise limit accuracy in parameter estimation~\footnote{The literature on quantum parameter estimation and metrological applications of many body systems is vast; {\it e.g.} see Refs.~[1-24] and references therein.}; actually, a first step in this direction using many-body systems has been recently realized: entangled states in systems of ultra-cold atoms
have been generated through spin-squeezing techniques~\cite{Oberthaler1,Treutlein}.
The aim is to use them as input states in interferometric apparatuses, specifically constructed
for quantum enhanced metrological applications.
In such devices, the initial $N$-qubit states are rotated by means of collective pseudo-spin operators; for distinguishable qubits, the relevance of entangled states is readily exposed by addressing single particle contributions to the collective operators~\cite{Sorensen,Briegel}; however, in the case of trapped ultra-cold atoms, the qubits involved are identical and thus not addressable, a fact that has often not been fully appreciated in the recent literature on quantum metrology and that has recently been tackled in Refs.~[25,26].

For $N$ distinguishable qubits, the notion of quantum entanglement is well-established (see the review~[27] and references therein) and strictly associated with the tensor product structure of the Hilbert space, $\mathbbm{H}_N=\bigotimes_{j=1}^N\mathbbm{H}_j$.
Such a structure is no longer available any longer for identical qubits and this asks for a different approach to the notions of non-locality and entanglement: it should not anymore refer  to the particle aspect of first quantization, rather be based on the behavior of correlation functions of commuting observables,
closely related to the mode description typical of second quantization.

In the following, we characterize the entangled states of $N$ Boson systems according to the generalized notion of separability given in Ref.~[25]; in particular, we show that the negativity, which measures the lack of positivity of partially transposed states, is an exhaustive bipartite entanglement witness.
This is to be contrasted with the case of distinguishable qubits where, apart for two qubits or one qubit and one qutrit, there exist entangled states with zero negativity~\cite{Hororev}.

Then, we show that a purely dephasing noise which, for distinguishable qubits, is responsible for mere decoherence, in the framework of identical Bosons can instead generate entanglement; however, we also show that this noise-induced entanglement cannot be used to improve on the sensitivities of matter interferometric devices based on such systems.

\section{Entangled Identical Bosons}

Since $N$-Boson density matrices must be convex combinations of symmetric $N$-particle states,
the tensor product structure which is natural for distinguishable particles is not appropriate
for discussing the entanglement properties of systems of identical particles.
Indeed, the Hilbert space for two identical Bosonic qubits is the sub-space of
$\mathbb{C}^4$ spanned by
$$
\vert \uparrow\uparrow\rangle\ ,\quad \vert \downarrow\downarrow\rangle\ ,\quad \frac{\vert \uparrow\downarrow\rangle+\vert\downarrow\uparrow\rangle}{\sqrt{2}}\ ,
$$
where $\vert i\rangle$, $i=\uparrow,\downarrow$, is any orthonormal basis in $\mathbb{C}^2$.

Entanglement should rather be investigated by means of creation $a_i^\dagger$
and annihilation operators $a_i$ of single-particle orthonormal basis states $\vert i\rangle$, obeying the canonical commutation relations $[a_i\,,\,a^\dag_j]=\delta_{ij}$.
Although the notion of entanglement in many-body systems has already
been addressed and discussed in the literature (for instance,
see Refs.~[28-44]), only limited results actually apply to the case of
identical particles. In Ref.~[25] it has been proposed that, when dealing with identical particles, entanglement should correspond to whether, given a state $\omega$ of the system, there are non-classical correlations among commuting observables belonging to two generic commuting sub-algebras $(\mathcal{A},\mathcal{B})$ of the entire algebra generated by the set of all creation and annihilation operators, which we shall refer to as an \textit{algebraic bipartition}.
An observable $\cal O$ is called ``$(\mathcal{A},\mathcal{B})$-local''
if it is of the form ${\cal O}=AB$, $A\in\mathcal{A}$ and
$B\in\mathcal{B}$, and a state $\omega$ ``$(\mathcal{A},\mathcal{B})$-separable''
if the expectation values $\omega(AB)$ of local operators $AB$ can be decomposed into a convex linear combination of product of expectations:
\begin{equation}
\label{def-sep}
\omega(AB)=
\sum_i\lambda_i\, \omega^a_i(A)\,\omega^b_i(B)\ ,\quad
\lambda_i>0\ ,\ \sum_i\lambda_i=1\ ,
\end{equation}
in terms of other states $\omega^{a}_i$, $\omega^{b}_i$;
otherwise, $\omega$ is said $(\mathcal{A},\mathcal{B})$-entangled.

The previous definition reduces to the standard notion of entanglement for distinguishable particles.
For instance, in the case of a two qubit system, by choosing the algebraic bipartition $\mathcal{A}=\mathcal{B}=M_2$, where $M_2$ is the algebra of $2\times 2$ matrices over $\mathbbm{C}^2$, and the expectation value $\omega(AB)={\rm Tr}\left(\rho\,A\otimes B\right)$ defined in the usual way
through the trace with the two-qubit density matrix $\rho$, condition~(\ref{def-sep}) readily gives that $\rho$ must be a convex combination of product states.

It should be stressed that, in the case of identical particles, there is no {\it a priori} given bipartition
so that questions about entanglement and separability, non-locality and locality
are meaningful only with reference to a specific class of (commuting) observables; the general definition given above precisely stems from this fundamental observation.

\subsection{Two Mode Boson Systems}
\label{sepAB-sec}

Consider a system of $N$ Bosons that can be found in only two different modes, like for instance a system consisting of ultracold atoms confined by a double-well potential.
Then, the states $\vert i\rangle$, $i=\uparrow,\downarrow$, describe one atom located within the left,
respectively the right well. These states correspond to two spatial modes and are created by the action on the vacuum state $\vert 0\rangle$ of creation operators $a^\dag$, $b^\dag$:
$a^\dag\vert0\rangle=\vert\!\downarrow\rangle$,
$b^\dag\vert0\rangle=\vert\!\uparrow\rangle$.

When the total number $N$ is conserved, the symmetric Fock space of this two-mode system
is generated by the $N+1$ orthonormal vectors:
\begin{equation}
\label{NFock}
\vert k,N-k\rangle_{\mathcal{A}\mathcal{B}}=\frac{(a^\dagger)^k (b^\dagger)^{N-k}}{\sqrt{k!(N-k)!}}\,
|0\rangle\ , \quad 0\leq k\leq N\ .
\end{equation}
Because of the orthogonality of the spatial modes, by considering the norm-closures of
all polynomials $P_a$ in $a,a^\dag$, respectively $P_b$ in $b,b^\dag$,
one obtains two commuting subalgebras $\mathcal{A}$ and $\mathcal{B}$ that in turn generate the whole algebra of bounded operators for this two mode Bosonic system.

In general, one can show that ($\mathcal{A},\mathcal{B}$)-separable density matrices
must be convex combinations of projections
$\vert k,N-k\rangle_{\mathcal{A}\mathcal{B}}\langle k,N-k\vert$ (see Ref.~[25]):
\begin{equation}
\label{sep}
\rho=\sum_{k=0}^N p_k \vert k,N-k\rangle_{\mathcal{A}\mathcal{B}}\langle k,N-k\vert\ , \qquad p_k>0\ ,\quad \sum_{k=0}^Np_k=1\ .
\end{equation}
This is so because $\mathcal{A}$ and $\mathcal{B}$ generate the whole two-mode $N$ Boson algebra
so that the only pure $(\mathcal{A},\mathcal{B}$)-separable states are projections onto the eigenstates~(\ref{NFock}) of the number operator $a^\dag a+b^\dag b$ which thus span the convex subset of $(\mathcal{A},\mathcal{B}$)-separable mixed states.

Notice that, in first quantization, an  ($\mathcal{A},\mathcal{B}$)-separable state for a
$N=2$ bosonic qubits like $a^\dag b^\dag\vert 0\rangle$ corresponds to
$(\vert\!\uparrow\downarrow\,\rangle+\vert\!\downarrow\uparrow\,\rangle)/\sqrt{2}$.
Such state is surely entangled for distinguishable qubits, while, according to our definition, it is no longer so
for identical bosonic qubits; the reason is that its entanglement is only formal as it comes from the necessary
symmetrization of the separable state $\vert\!\uparrow\downarrow\rangle$~\cite{Ghirardi2002}.

For bipartite systems of distinguishable particles one knows that states $\rho$ that do not remain positive under partial transposition are entangled~\cite{Hororev} and are witnessed by the so-called {\it negativity}:
\begin{equation}
\label{negativity}
\mathcal{N}(\rho)=\|\rho^\Gamma\|_1-1\ ,\qquad \|\rho^\Gamma\|_1={\rm Tr}\Big(\sqrt{(\rho^\Gamma)^\dag\rho^\Gamma}\,\Big)\ ,
\end{equation}
where $\rho^\Gamma$ is the transposition with respect to the first party.

Since ${\rm Tr}(\rho)={\rm tr}(\rho^\Gamma)=1$, if $\rho^\Gamma$ does not remain positive
under partial transposition, then $\|\rho^\Gamma\|_1>1$ and $\mathcal{N}(\rho)>0$.
Unfortunately, there can be entangled states that remain positive under partial transposition whence $\mathcal{N}(\rho)=0$; therefore, the negativity is not
an exhaustive entanglement witness for generic bipartite states of distinguishable particles.

Remarkably, negativity is instead an exhaustive entanglement witness for the case at hands: indeed, by performing the partial transposition with respect to the first mode of a generic two mode $N$-Boson state
\begin{equation}
\label{NbosDM}
\rho=\sum_{k,\ell=0}^N\rho_{k\ell}\vert k,N-k\rangle_{\mathcal{A}\mathcal{B}}\langle \ell,N-\ell\vert\ ,\qquad\sum_{k=0}^N\rho_{kk}=1\ ,
\end{equation}
one obtains an operator on a larger Hilbert space than the sector of the Fock space with fixed $N$, namely
$$
\rho^\Gamma=\sum_{k,\ell=0}^N\rho_{k\ell}\vert \ell,N-k\rangle_{\mathcal{A}\mathcal{B}}\langle k,N-\ell\vert\ ,
$$
which is such that
$$
(\rho^\Gamma)^\dag\rho^\Gamma=\sum_{k,\ell}|\rho_{k\ell}|^2\,\vert k,N-\ell\rangle_{\mathcal{A}\mathcal{B}}\langle k,N-\ell\vert\ ,
$$
whence the negativity
\begin{equation}
\label{Negativity1}
\mathcal{N}(\rho)=\sum_{k\neq\ell=0}^N\,\Big|\rho_{k\ell}\Big|
\end{equation}
vanishes if and only if $\rho$ has null off-diagonal element with respect to the Fock states relative to the chosen bipartition, {\it i.e.} it is separable because of the form~(\ref{sep}).

\subsection{Bogolubov Transformations and Entanglement}

While the Fock number states~(\ref{NFock}) are $(\mathcal{A},\mathcal{B})$-separable, important examples of $(\mathcal{A},\mathcal{B})$-entangled states are the so-called
discrete coherent states
\begin{eqnarray}
\label{coherent0a}
\vert\xi,\varphi\rangle_{\mathcal{A}\mathcal{B}}&=&\frac{1}{\sqrt{N!}}\,
\Big(\sqrt{\xi}{\rm e}^{-i\varphi/2}\, a^\dag\,+\,
\sqrt{1-\xi}{\rm e}^{i\varphi/2}\,b^\dag\Big)^N\,\vert 0\rangle\\
&=&\sum_{k=0}^N\sqrt{{N\choose k}}\xi^{k/2}(1-\xi)^{(N-k)/2}\,{\rm e}^{-ik\varphi+iN\varphi/2}\,
\vert k,N-k\rangle_{\mathcal{A}\mathcal{B}}\ ,
\end{eqnarray}
where $0\leq\xi\leq 1$. These states describe the situation when all $N$ Boson are in the same single particle state
$(\sqrt{\xi}\exp(-i\varphi/2)\,,\,\sqrt{1-\xi}\exp(i\varphi/2))$: their off-diagonal elements do not vanish.
 Therefore, the corresponding negativity is also non-vanishing; indeed, it reads
\begin{equation}
\label{negcoh}
\mathcal{N}\Big(\vert\xi,\varphi\rangle_{\mathcal{A}\mathcal{B}}\langle\xi,\varphi\vert\Big)=\sum_{k\neq\ell}\sqrt{{N\choose k}{N\choose\ell}}\xi^{(k+\ell)/2}(1-\xi)^{N-(k+\ell)/2}\ .
\end{equation}

The Bogolubov transformation
\begin{equation}
\label{bogol}
c=\frac{a+b}{\sqrt{2}}\ ,\quad d=\frac{a-b}{\sqrt{2}}\ ,
\end{equation}
changes the spatial modes $a,b$ into energy modes; indeed, it corresponds to a change of basis from that of spatially localized states, to the one of the eigenstates
$c^\dag\vert0\rangle=(\vert\!\downarrow\rangle+\vert\!\uparrow\rangle)/2$,
$d^\dag\vert0\rangle=(\vert\!\downarrow\rangle-\vert\!\uparrow\rangle)/2$
of a single particle Hamiltonian with a highly penetrable barrier.

The algebras $\mathcal{C}$, respectively $\mathcal{D}$ constructed by means of polynomials in $c,c^\dag$, respectively $d,d^\dag$ commute, generate the two-mode $N$ Boson algebra and thus provide another possible algebraic bipartition. It thus turn out that the $(\mathcal{A},\mathcal{B})$-entangled coherent state
\begin{equation}
\label{coherent0b}
\vert 1/2,0\rangle_{\mathcal{A}\mathcal{B}}=\frac{1}{\sqrt{N!}}\Big(\frac{a^\dag\,+\,b^\dag}{\sqrt{2}}\Big)^N
=\frac{(c^\dag)^N}{\sqrt{N!}}\vert0\rangle\ ,
\end{equation}
results the Fock number state $\vert N,0\rangle_{\mathcal{C}\mathcal{D}}$ for the
number operator $c^\dag\,c+d^\dag\,d$, and therefore it results  $(\mathcal{C},\mathcal{D})$-separable.
Thus, when we refer to the negativity of a state $\rho$, one must specify with respect to which algebraic bipartition the partial transposition is performed; indeed,
$$
\mathcal{N}_{\mathcal{A}\mathcal{B}}\Big(\vert 1/2,0\rangle_{\mathcal{A}\mathcal{B}}\langle1/2,0\vert\Big)>0\ ,\quad\hbox{while}\quad
\mathcal{N}_{\mathcal{C}\mathcal{D}}\Big(\vert 1/2,0\rangle_{\mathcal{A}\mathcal{B}}\langle1/2,0\vert\Big)=0\ .
$$

As another instance of the dependence of the notions of entanglement and non-locality on the chosen bipartition, consider the operators
\begin{equation}
\label{jsAB}
J_x=\frac{1}{2}(a^\dag b+ab^\dag)\ ,\
J_y=\frac{1}{2i}(a^\dag b-ab^\dag)\ ,\
J_z=\frac{1}{2}(a^\dag a-b^\dag b)\ ,
\end{equation}
that satisfy the $SU(2)$ algebraic relations
$[J_x\,,\,J_y]=i\,J_z$ and their cyclic permutations.
They are all non-local with respect to the algebraic bipartition $(\mathcal{A},\mathcal{B})$
and such are the rotations
$\displaystyle
{\rm e}^{i\theta J_x}$ and $\displaystyle{\rm e}^{i\theta J_y}$ they generate,
while
$\displaystyle{\rm e}^{i\theta J_z}={\rm e}^{i\theta a^\dag a}{\rm e}^{-i\theta b^\dag b}$
is $(\mathcal{A},\mathcal{B})$-local.

By means of the Bogolubov transformation~(\ref{bogol}) one rewrites
$$
J_x=\frac{1}{2}(c^\dag c-d^\dag d)\ ,\
J_y=\frac{1}{2i}(d^\dag c-dc^\dag)\ ,\
J_z=\frac{1}{2}(c^\dag d+ cd^\dag)\ .
$$
Relatively to $(\mathcal{C},\mathcal{D})$, it is now
$\displaystyle{\rm e}^{i\theta\, J_x}={\rm e}^{i\theta\, c^\dag c}{\rm e}^{-i\theta\, d^\dag d}$
which acts locally.

\section{Two-mode $N$ Bosons and Noise}

An important feature of matter interferometry based upon ultracold atoms trapped in double well potential is the coherence between the spatial modes; this is endangered by the presence of a dephasing noise that tends to suppress the off-diagonal matrix elements
$\rho_{k\ell}={}_{\mathcal{A}\mathcal{B}}\langle k,N-k\vert\rho\vert\ell,N-\ell\rangle_{\mathcal{A}\mathcal{B}}$, $k\neq\ell$,
with respect to the orthonormal basis of Fock states~(\ref{NFock}). The effects of this kind of noise can be described by the following master equation~\footnote{This master equation is standard in the theory of open quantum systems;
for details see \hbox{Refs.~[45-49]}. For more recent applications to trapped ultracold gases, {\it e.g.} see Ref.~[50-55].}
\begin{equation}
\label{dephasing}
\partial_t\rho(t)=\gamma\Big(J_z\,\rho(t)\,J_z\,-\,\frac{1}{2}\{J_z^2\,,\,\rho(t)\}\Big)\ ,
\end{equation}
where $\rho$ is the $N$ Boson density matrix, $\gamma$ measures the strength of the noise and $J_z$
is the collective spin operator in~(\ref{jsAB}) that commutes with the number operator $a^\dag\,a+b^\dag\,b$.
One easily checks that the matrix elements with respect to the eigenstates~(\ref{NFock}) satisfy
\begin{equation}
\label{mastereq}
\partial_t\rho_{k\ell}(t)=-\frac{\gamma}{2}(k-\ell)^2\,\rho_{k\ell}\ ,
\end{equation}
whence this kind of noisy irreversible time-evolution tends to diagonalize the system states with respect to their basis:
\begin{equation}
\label{sol1}
\rho(t)=\sum_{k\ell=0}^N{\rm e}^{-t\gamma(k-\ell)^2/2}\,\rho_{k\ell}\,
\vert k,N-k\rangle_{\mathcal{A}\mathcal{B}}\langle\ell,N-\ell\vert\ .
\end{equation}
It thus follows that the $(\mathcal{A},\mathcal{B})$-negativity either decreases exponentially
in time,
\begin{equation}
\label{negnoise}
\mathcal{N}_{\mathcal{A},\mathcal{B}}(\rho(t))=\sum_{k\neq\ell}{\rm e}^{-t\gamma(k-\ell)^2/2}\,|\rho_{k\ell}|\leq {\rm e}^{-t\gamma/2}\,\mathcal{N}_{\mathcal{A},\mathcal{B}}(\rho)\ ,
\end{equation}
if the initial state is $(\mathcal{A},\mathcal{B})$-entangled or remains zero as the dephasing noise
has no possibility of creating non-local effects with respect to the bipartition $(\mathcal{A},\mathcal{B})$.
This is best seen by rewriting the solution~(\ref{sol1}) in the more suggestive form
\begin{equation}
\label{sol2}
\rho(t)=\frac{1}{2\sqrt{\pi}}\int_{-\infty}^{+\infty}{\rm d}u\,{\rm e}^{-u^2/4}\,
{\rm e}^{-i\sqrt{t\gamma/2}\,u\,J_z}\,\rho\,{\rm e}^{+i\sqrt{t\gamma/2}\,u\,J_z}\ ,
\end{equation}
which, on one hand, explicitly exhibits the Kraus from of the completely positive maps solutions to~(\ref{mastereq}) and, on the other hand, shows the impossibility of generating $(\mathcal{A},\mathcal{B})$-entanglement as the rotations generated by $J_z$ are all  $(\mathcal{A},\mathcal{B})$-local.

However, they are not local with respect to the bipartition $(\mathcal{C},\mathcal{D})$
obtained by the Bogolubov transformation~(\ref{bogol}) and are thus able to raise from zero the
$(\mathcal{C},\mathcal{D})$-negativity of an initial $(\mathcal{C},\mathcal{D})$-separable state.
For instance, consider the pure state~(\ref{coherent0b}) as initial state.
Using~(\ref{bogol}) one finds
$$
{\rm e}^{-i\beta J_z}\vert N,0\rangle_{\mathcal{C}\mathcal{D}}=\frac{1}{\sqrt{N!}}\Big(\frac{a^\dag{\rm e}^{-i\beta/2}+
b^\dag{\rm e}^{i\beta/2}}{\sqrt{2}}\Big)^N\vert0\rangle=
\Big\vert\cos^2\beta/2,\pi/2\Big\rangle_{\mathcal{C}\mathcal{D}}\ ,
$$
so that the time-evolving state reads
\begin{equation}
\label{exat}
\rho(t)=\frac{1}{2\sqrt{\pi}}\int_{-\infty}^{+\infty}{\rm d}u\,{\rm e}^{-u^2/4}\,
\Big\vert\cos^2(u\sqrt{t\gamma/2}),\pi/2\Big\rangle_{\mathcal{C}\mathcal{D}}
\Big\langle\cos^2(u\sqrt{t\gamma/2}),\pi/2\Big\vert\ .
\end{equation}
The time-evolution thus results in a mixed state which surely has non-vanishing off-diagonal elements
\begin{eqnarray*}
&&\hskip-1cm
{}_{\mathcal{C}\mathcal{D}}\langle k,N-k|\rho(t)\vert \ell,N-\ell\rangle_{\mathcal{C}\mathcal{D}}=
\frac{1}{\sqrt{\pi}}\int_0^{+\infty}{\rm d}u\,{\rm e}^{-u^2/4}\,
\sqrt{{N\choose k}{N\choose\ell}}\ \times\\
&&\hskip 1cm
\times\
\cos^{k+\ell} \Big(u\sqrt{\frac{t\gamma}{2}}\Big)\,\sin^{2N-k-\ell}\Big(u\sqrt{\frac{t\gamma}{2}}\Big)
\,{\rm e}^{-i\pi/4(k-\ell)}\ ,
\end{eqnarray*}
with respect to the orthonormal basis of eigenstates of $c^\dag c+d^\dag d$ and thus $\mathcal{N}_{\mathcal{C}\mathcal{D}}(\rho(t))>0$.

\subsection{Entanglement of Identical Particles and Noise in Metrological Appplications}

The use of $N$ Bosons states for measuring an angle $\theta$ by interferometric techniques is based on the following scheme: an input state $\rho$ is rotated into a final state
\begin{equation}
\label{state-rot}
\rho_\theta={\rm e}^{-i\theta J_{\vec{n}_1}}\,\rho\,{\rm e}^{i\theta J_{\vec{n}_1}}\ ,
\end{equation}
where $\theta$ by means of the collective spin operator in the direction of a unit vector $\vec{n}_1$: $J_{\vec{n}_1}=\vec{n}_1\cdot\vec{J}$; then, on the rotated state one performs the measurement of a collective spin operator $J_{\vec{n}_2}$, where $\vec{n}_2\perp\vec{n}_1$ and, for sake of convenience, we choose $\rho$ such that $\langle J_{\vec{n}_3}\rangle={\rm Tr}(\rho\, J_{\vec{n}_3})\neq 0$ along the third orthogonal unit vector $\vec{n}_3$. By error propagation, the mean square error $\Delta J_{\vec{n}_2}=\sqrt{\langle J_{\vec{n}_2}^2\rangle-\langle J_{\vec{n}_2}\rangle^2}$ is related to the error $\delta\theta$ in the measurement of the small rotation angle $\theta$ by~\cite{Wineland}:
\begin{equation}
\label{Squeezparam0}
\delta^2\theta=\frac{\Delta^2 J_{\vec{n}_2}}{\Big(\partial_\theta\langle
J_{\vec{n}_2}\rangle_\theta\Big\vert_{\theta=0}\Big)^2}
=\frac{\Delta^2 J_{\vec{n}_2}}{\langle J_{\vec{n}_3}\rangle^2}=\frac{\xi_W^2}{N}\ ,
\end{equation}
where the parameter
\begin{equation}
\label{Squeezparam}
\xi_W^2:=\frac{N\Delta^2 J_{\vec{n}_2}}{\langle J_{\vec{n}_3}\rangle^2}\ .
\end{equation}
measure the amount of squeezing in the state $\rho$. Indeed, for any orthogonal triplet of space-directions $\vec{n}_1$, $\vec{n}_2$, $\vec{n}_3$,
the Heisenberg uncertainty relations read
\begin{equation}
\Delta^2 J_{\vec{n}_1}\Delta^2 J_{\vec{n}_2}\geqslant\frac{1}{4}\langle J_{\vec{n}_3}\rangle^2 \ ,
\end{equation}
and $\rho$ is a squeezed state if one of the variances can be made smaller than
$\frac{1}{2}\,\Big|\langle J_{\vec{n}_3}\rangle\Big|$.

The value $\displaystyle\delta^2\theta=1/N$ is called shot-noise limit; in the case of
distinguishable qubits, it gives the lower bound to the attainable accuracies when the input
state $\rho$ is separable~\cite{Sorensen}. Therefore, for systems consisting of distinguishable qubit, entanglement in the initial state $\rho$ is necessary to achieve sub-shot-noise accuracies. An entangled initial state is usually prepared by preliminary squeezing operations~\cite{Oberthaler1,Treutlein}; indeed, from~(\ref{Squeezparam0}), preparing the initial state $\rho$ such that $\xi_W^2<1$ guarantees an achievable sub-shot-noise accuracy in the determination of $\theta$. However, in Ref.~[26] it is shown that this is not strictly necessary in ultracold atom interferometry; indeed, in such experimental contexts, one is dealing with identical Bosons and then the necessary non-local effects necessary for beating the shot-noise limit can be provided  by the interferometric apparatus itself.

In the previous section we have seen that the presence of dephasing noise destroys $(\mathcal{A}\mathcal{B})$-entanglement, but creates $(\mathcal{C}\mathcal{D})$-entanglement.
It is thus of interest to see whether this latter fact allows one to achieve sub-shot-noise accuracies by simply letting the $(\mathcal{C}\mathcal{D})$-non-local noisy mechanism act.
In order to do so we need compute the mean values $\langle J_{\vec{n}}\rangle_t$ and
$\langle J^2_{\vec{n}}\rangle_t$ of collective spin operator aligned along the unit vector $\vec{n}$ with respect to the time-evolving state~(\ref{sol1}); use of~(\ref{sol2}) yields
\begin{equation}
\label{mean}
\langle J_{\vec{n}}\rangle_t=\frac{1}{2\sqrt{\pi}}\int_{-\infty}^{+\infty}{\rm d}u\,{\rm e}^{-u^2/4}\,
{\rm Tr}(\rho\,J_{\vec{n}(u,t)})\ ,
\end{equation}
in terms of the rotated unit vector
\begin{equation}
\label{unitv}
\vec{n}(u,t)=\begin{pmatrix}\cos\Big(u\sqrt{\frac{t\gamma}{2}}\Big)&
\sin\Big(u\sqrt{\frac{t\gamma}{2}}\Big)&0\cr
-\sin\Big(u\sqrt{\frac{t\gamma}{2}}\Big)&
\cos\Big(u\sqrt{\frac{t\gamma}{2}}\Big)&0\cr
0&0&1\end{pmatrix}\,\vec{n}\ .
\end{equation}
Furthermore, given the mean value
\begin{equation}
\label{variance}
\langle J^2_{\vec{n}}\rangle_t=\frac{1}{2\sqrt{\pi}}\int_{-\infty}^{+\infty}{\rm d}u\,{\rm e}^{-u^2/4}\, {\rm Tr}(\rho\,J^2_{\vec{n}(u,t)})\ ,
\end{equation}\\
one finds
\begin{eqnarray*}
&&
\Delta^2_tJ_{\vec{n}}=\langle J^2_{\vec{n}}\rangle_t-\langle J_{\vec{n}}\rangle_t^2=\frac{1}{2\sqrt{\pi}}\int_{-\infty}^{+\infty}{\rm d}u\,{\rm e}^{-u^2/4}\, \Delta^2J_{\vec{n}(u,t)}\ +\\
&&
+\frac{1}{2\sqrt{\pi}}\int_{-\infty}^{+\infty}{\rm d}u\,{\rm e}^{-u^2/4}\,\Big({\rm Tr}(\rho\,J_{\vec{n}(u,t)})\Big)^2\,-\,\Bigg(\frac{1}{2\sqrt{\pi}}\int_{-\infty}^{+\infty}{\rm d}u\,{\rm e}^{-u^2/4}\,{\rm Tr}(\rho\,J_{\vec{n}(u,t)})\Bigg)^2\ .
\end{eqnarray*}
Because of the convexity of the function $f(x)=x^2$, the second line above is positive and one estimates
$$
\Delta^2_tJ_{\vec{n}}=\langle J^2_{\vec{n}}\rangle_t-\langle J_{\vec{n}}\rangle_t^2\,\geq\,
\frac{1}{2\sqrt{\pi}}\int_{-\infty}^{+\infty}{\rm d}u\,{\rm e}^{-u^2/4}\, \Delta^2J_{\vec{n}(u,t)}\ .
$$
If the initial state $\rho$ is such that for no orthogonal directions $\vec{n}_{2,3}$ the squeezing parameter~(\ref{Squeezparam}) is less than one, then, as $\vec{n}_2(u,t)\perp\vec{n}_3(u,t)$, one gets
$$
\frac{1}{2\sqrt{\pi}}\int_{-\infty}^{+\infty}{\rm d}u\,{\rm e}^{-u^2/4}\, \Delta^2J_{\vec{n}_2(u,t)}\geq
\frac{1}{N}\frac{1}{2\sqrt{\pi}}\int_{-\infty}^{+\infty}{\rm d}u\,{\rm e}^{-u^2/4}\, \langle J_{\vec{n}_3(u,t)}\rangle^2\ ,
$$
and, again by convexity,
\begin{eqnarray*}
\frac{1}{N}\frac{1}{2\sqrt{\pi}}\int_{-\infty}^{+\infty}{\rm d}u\,{\rm e}^{-u^2/4}\, \langle J_{\vec{n}_3(u,t)}\rangle^2
&\geq&
\frac{1}{N}\Bigg(\frac{1}{2\sqrt{\pi}}\int_{-\infty}^{+\infty}{\rm d}u\,{\rm e}^{-u^2/4}\, \langle J_{\vec{n}_3(u,t)}\rangle\Bigg)^2\\
&=&\frac{1}{N}\langle J_{\vec{n}_3}\rangle_t^2\ .
\end{eqnarray*}
Thus, although under the dephasing noise some $N$-Bosons states can get entangled, no
squeezing can be achieved by such means; indeed,
\begin{equation}
\label{squeez}
\frac{N\Delta^2_tJ_{\vec{n}_2}}{\langle J_{\vec{n}_3}\rangle_t^2}\geq 1\ .
\end{equation}
Though not metrologically relevant, such an environment generated entanglement might however be useful for other quantum informational tasks like those involving quantum gates constructed by using systems of ultracold Bosons trapped in optical lattices~\cite{Bloch}.

\section{Conclusions}

In the case of systems of identical Bosons, it is necessary to introduce a new, algebraic definition of separability; indeed, the standard one based on an a priori factorized form of the involved Hilbert space is no longer available. Based on a natural notion of entanglement given in terms of correlations among commuting sub-algebras of observables, we have first proved that, unlike for distinguishable qubits, entangled states of $N$ Bosonic qubits are completely identified by their non zero negativity which is therefore an exhaustive bipartite entanglement witness for such systems.
Furthermore, we have showed that even a simple dephasing noise which, for distinguishable particles exhibits merely decoherence effects, can instead generate entanglement among identical Bosons.
These results may be relevant in concrete applications to systems of ultracold atoms trapped in optical lattices; however, while the entanglement generated by purely dephasing noise can be used for the practical implementation of quantum informational protocols, it cannot augment the sensitivity of ultracold atom based interferometric devices.

\end{document}